\documentclass[useAMS,usenatbib]{mn2e}
\usepackage{txfonts}
\usepackage{natbib}
\usepackage[all]{xy}
\usepackage[dvips]{graphicx}

\def\del#1{{}}

\newcommand{\ltsima}{$\; \buildrel < \over \sim \;$}
\newcommand{\lsim}{\lower.5ex\hbox{\ltsima}}
\newcommand{\gtsima}{$\; \buildrel > \over \sim \;$}
\newcommand{\gsim}{\lower.5ex\hbox{\gtsima}}
\newcommand{\bra}{\langle}
\newcommand{\ket}{\rangle}

\newcommand{\dd}{\mathrm{d}}
\newcommand{\e}{\mathrm{e}}
\newcommand{\p}{\mathrm{p}}

\newcommand{\trace}{\mathrm{tr}}

\newcommand{\dirac}{\delta_\mathrm{D}}

\newcommand{\be}{\begin{equation}}
\newcommand{\ee}{\end{equation}}

\title[iSW-effect with orthogonal polynomials]
{Integrated Sachs-Wolfe tomography with orthogonal polynomials}
\author[Gero J{\"u}rgens and B.M. Sch{\"a}fer]{
Gero J{\"u}rgens\thanks{gero.juergens@stud.uni-heidelberg.de}$^1$ and Bj{\"o}rn Malte Sch{\"a}fer$^2$\\
$^1$Institut f{\"u}r theoretische Astrophysik, Zentrum f{\"u}r Astronomie, Universit{\"a}t Heidelberg, Albert-Ueberle-Stra{\ss}e 2, 69120 Heidelberg, 
Germany\\$^2$Astronomisches Recheninstitut, Zentrum f{\"u}r Astronomie, Universit{\"a}t Heidelberg, M{\"o}nchhofstra{\ss}e 12, 69120 Heidelberg, Germany}

\begin{document}
\pagerange{\pageref{firstpage}--\pageref{lastpage}}
\pubyear{2008}
\maketitle
\label{firstpage}

\begin{abstract}
Topic of this article are tomographic measurements of the integrated Sachs-Wolfe effect with specifically designed, 
orthogonal polynomials which project out statistically independent modes of the galaxy distribution. The polynomials are contructed using the 
Gram-Schmidt orthogonalisation method.
To quantify the power of the iSW-effect in contraining cosmological parameters we perfom a combined Fisher matrix analysis for the iSW-, galaxy- and 
cross-spectra for $w$CDM cosmologies using the survey characteristics of PLANCK and EUCLID. The signal to noise ratio has also been studied for other 
contemporary galaxy surveys, such as SDSS, NVSS and 2MASS. For the cross-spectra our tomographic method provides a $16\%$ increase in the signal to noise ratio and an improvement of up to $30\%$ in conditional errors 
on parameters. Including all spectra, the marginalised errors approach an inverse square-root dependence with increasing cumulative polynomial order which 
underlines the statistical independence of the weighted signal spectra. 
\end{abstract}


\section{Introduction}
The integrated Sachs-Wolfe (iSW) effect is one of the secondary anisotropies of the cosmic microwave background (CMB). Time-evolving gravitational 
potentials in the large-scale structure generate temperature fluctuations in the CMB \citep{Sachs1967}. The iSW-effect is a valuable tool for investigating 
dark energy and non-standard cosmologies since it is sensitive to fluids with non-zero equation of state \citep{Crittenden1996}. 
For this reason its detection is of particular relevance for cosmology and the nature of gravity \citep{Lue2004,Zhang2006} even if its signal strength is very low.

Since the iSW-effect is generated in time-evolving potential wells for photons on their way from the last scattering surface to us, it will be strongly 
correlated with the galaxy density field. Therefore, the cross-spectrum will provide valuable additional cosmological information. The iSW 
effect has been measured in such cross-correlation studies \citep{Boughn1998,Boughn2004,Vielva2006,mcewen2007,Giannantonio2008}.
However, due to the line of sight integration, a detailed distance resolution of the processes can not be withdrawn from these spectra.

A former approach correlated large scale structure observations from various survey with the CMB anisotropies to study the iSW-effect as a function of redshift and 
to formulate a reliable likelihood formulation for parameter constraints \citep{Ho2008}. Also recently, \cite{Frommert2008} presented an optimal method to 
reduce the local variance effect and gained 7 per cent in the signal to noise ratio for the cross-spectra.

In this work we aim to formulate a tomographic approach with help of an orthogonal set of weighting polynomials, 
which is similar to a former application to weak lensing spectra \citep{Schaefer2011}. 
The orthogonality of the polynomials will generically lead to a diagonal signal covariance matrix and will 
therefore provide cumulative statistical independent measurements with increasing polynomial order.

The article has the following structure: In Section~\ref{sect_foundations} we provide introductory information about dark energy cosmologies, 
CDM power spectra and linear structure growth within these cosmologies (Sections~\ref{sect_de_cosmo}-\ref{sect_structure_growth}). 
We also introduce a galaxy distribution function (Section~\ref{sect_galaxy_distribution}) and give a short introduction to the iSW-effect 
(Section~\ref{sect_iSW}).
The orthogonal polynomials are motivated and constructed in Section~\ref{sect_motivation} and \ref{sect_construction}, also their most important 
properties  are discussed (Section~\ref{sect_properties}). In Section~\ref{sect_statistics} we discuss how tomography with orthogonal polynomials 
can improve statistical constraints on cosmological parameters. After calculating the noise contributions (Section~\ref{sect_covariance}) we perform a 
Fisher matrix analysis (Section~\ref{sect_fisher}) and discuss signal to noise ratios and statistical errors (Section~\ref{sect_s2n}-\ref{sect_errors}).
The results are summarised in Section~\ref{sect_summary}. 

The reference cosmological model used is a spatially flat $w$CDM cosmology with Gaussian adiabatic initial perturbations 
in the cold dark matter density field. The specific parameter choices are 
$\Omega_{\mathrm m} = 0.25$, $n_{\mathrm s} = 1$, $\sigma_8 = 0.8$, $\Omega_\mathrm{b}=0.04$ and $H_0=100\: h\:\mathrm{km}/\mathrm{s}/\mathrm{Mpc}$, with $h=0.72$. 
The dark energy equation of state is set to $w=-0.9$ and the sound speed is equal to the speed of light, $c_\mathrm{s}=c$.
\section{Cosmology and  iSW-effect}\label{sect_foundations}

\subsection{Dark energy cosmologies}\label{sect_de_cosmo}
In spatially flat dark energy cosmologies with the matter density parameter $\Omega_{\mathrm m}$, the Hubble function $H(a)=\dd\ln a/\dd t$ is given by
\begin{equation}
\frac{H^{\,2}(a)}{H_0^{\,2}} = \Omega_{\mathrm m}\,a^{\,-3} + (1-\Omega_{\mathrm m})\,a^{\,-3\,(1+w)},
\end{equation}
with a constant dark energy equation of state parameter $w$. The value $w\equiv -1$ corresponds to the cosmological constant $\Lambda$. 
The relation between comoving distance $\chi$ and scale factor $a$ is given by
\begin{equation}
\chi = c\int_a^1\dd a\:\frac{1}{a^2 H(a)},
\end{equation}
in units of the Hubble distance $\chi_H=c/H_0$.

\subsection{CDM power spectrum}\label{sect_cdm_ps}
The linear CDM density power spectrum $P(k)$ describes the fluctuation amplitude of the Gaussian homogeneous density field 
$\delta$, $\bra\delta(\bmath{k})\delta^*(\bmath{k}^\prime)\ket=(2\pi)^3\dirac(\bmath{k}-\bmath{k}^\prime)P(k)$, and is given by the ansatz
\begin{equation}
P(k)\propto k^{\,n_{\mathrm s}}T^2(k),
\end{equation}
with the transfer function $T(k)$. In low-$\Omega_{\mathrm m}$ cosmologies $T(k)$ is approximated with the fit proposed by \cite{Bardeen1986},
\begin{eqnarray}
T(q) &= &\frac{\ln(1+2.34\,q)}{2.34\,q}\nonumber \\
   & &\times \,\left[1+3.89\,q+(16.1\,q)^2+(5.46\,q)^3+(6.71\,q)^4\right]^{-1/4}\hspace{0.1 cm},
\label{eqn_cdm_transfer}
\end{eqnarray}
where the wave number $k=q\,\Gamma$ is rescaled with the shape parameter $\Gamma$ \citep{Sugiyama1995} 
which assumes corrections due to the baryon density $\Omega_\mathrm{b}$,
\begin{equation}
\Gamma=\Omega_{\mathrm m} h\,\exp\left[-\Omega_\mathrm{b}\left(1+\frac{\sqrt{2h}}{\Omega_{\mathrm m}}\right)\right].
\end{equation}
The spectrum $P(k)$ is normalised to the variance $\sigma_8$ on the scale $R=8~\mathrm{Mpc}/h$,
\begin{equation}
\sigma^2_R 
= \frac{1}{2\pi^2}\int\dd k\:k^2 P(k) W^2(kR)
\end{equation}
with a Fourier transformed spherical top hat filter function, $W(x)=3j_1(x)/x$. $j_\ell(x)$ is the 
spherical Bessel function of the first kind of order $\ell$  \citep{Abramowitz1972}.

\subsection{Structure growth with clustering dark energy}\label{sect_structure_growth}
Linear homogeneous growth of the density field, $\delta(\bmath{x},a)=D_+(a)\,\delta(\bmath{x},a=1)$, is described by the growth function $D_+(a)$, 
which is the solution to the growth equation \citep{Turner1997,Wang1998,Linder2003},
\begin{equation}
\frac{\dd^2}{\dd a^2}D_+(a) + \frac{1}{a}\left(3+\frac{\dd\ln H}{\dd\ln a}\right)\frac{\dd}{\dd a}D_+(a) = 
\frac{3}{2\,a^2}\,\Omega_{\mathrm m}(a) \,D_+(a) \hspace{0.1 cm}.
\label{eqn_growth}
\end{equation}

\subsection{Galaxy distribution}\label{sect_galaxy_distribution}
Galaxies form when strong peaks in the density field decouple from the Hubble expansion due to self-gravity. These so called protohalos approximately 
undergo an elliptical collapse \citep{Mo2007,Sheth2001}. 

In contrary to the pressureless dark matter component the baryons inside a 
dark matter halo can loose energy via radiative cooling and form stars. Because of this different behaviour, strictly speaking, 
one can not deduce the fractional perturbation $\Delta n/\bra n\ket$ in the mean number 
density of galaxies $\bra n \ket$ from the dark matter overdensity $\delta=\Delta \rho/\rho$. 
In a very simple way, however, the linear relation between the two entities,
\be
 \frac{\Delta n}{\bra n \ket} = b\,\frac{\Delta\rho}{\bra \rho \ket}\hspace{0.1 cm},
\ee
is a good approximation in most cases and was proposed by \cite{Bardeen1986}. 
The bias parameter $b$ can generally depend on scale \citep{Lumsden1989}, time \citep{Fry1996,Tegmark1998} 
as well as the galaxies luminosity and morphology. For simplicity we set the 
galaxy bias to unity throughout this paper, $b\equiv1$.
An established parametrisation of the redshift distribution $n(z)\,dz$ of galaxies is
\be
 n(z)\,dz = n_0\, \left(\frac z{z_0}\right)^2\, \exp \left[ -\left(\frac z{z_0}\right)^\beta \right]\ dz \hspace{0.3 cm}\hbox{with}\hspace{0.3 cm}
 \frac {1}{n_0} = \frac{z_0}{\beta}\, \Gamma\left( \frac 3\beta\right)
\ee
wich was introduced by \cite{Smail1995} and will also be used in this work. The parameter $z_0$ is related to the median redshift of the galaxy sample 
$z_{\mathrm{med}}=1.406\, z_0$ if $\beta = 3/2$. Finally, the $\Gamma$-function \citep{Abramowitz1972} determines the normalisation parameter $n_0$.  

\subsection{The integrated Sachse-Wolfe (iSW) effect}\label{sect_iSW}
Due to its expansion our universe had cooled down sufficiently to allow the formation of hydrogen atoms at a redshift of $z \simeq 1089$ 
\citep{Spergel2003}. Fluctuations in the gravitational potential imposed a shift in the decoupled photons which were emitted in the 
(re)combination process (Sachse-Wolfe effect). This primary anisotropy can be observed in the cosmic microwave background (CMB) 
in form of temperature fluctuations $\Delta T/T_{\mathrm{CMB}} \simeq 10^{-5}$ on large scales around its mean temperature 
$T_{\mathrm{CMB}}=2.726$ K \citep{Fixsen2009}.

Besides this, photons are subjected to several other effects on their way to us, which lead to secondary anisotropies \citep{Aghanim2008}, 
of which only the most important ones are mentioned here: 
Gravitational lensing \citep{Bartelmann2001}, Compton-collisions with free cluster electrons \citep[Sunyaev-Zel´dovich effect,][]{Zeldovich1980} 
and with electrons in uncollapsed structures  \citep[Ostriker-Vishniac effect,][]{Ostriker1986} and gravitational coupling to linear time-evolving potential wells 
\citep[integrated Sachs-Wolfe effect,][ which will be subject of this work]{Sachs1967}.

Assuming a completely transparent space, i.e. vanishing optical depth due to compton scattering $\tau_{\mathrm{opt}}(\eta)=0$, 
the temperature fluctuations $\tau(\hat\theta)$ generated by the iSW-effect can be expressed by the line of sight integral \citep{Sachs1967} 
\be
 \tau(\bmath\theta) \equiv \frac {\Delta T_{\mathrm{iSW}}}{T_{\mathrm{CMB}}} = \frac{2}{c^3} \, \int_0^{\chi_H} \dd\chi\, 
  a^2\, H(a)\,\frac{\partial}{\partial a} \Phi\,(\bmath\theta \chi, \chi)\hspace{0.1 cm} ,
\ee
reaching out to the limit of Newtonian gravity. Using the Poisson equation we can write this integral in terms of the dimensionless 
potential $\phi = \Phi/ \chi_{H}^2 = \Delta^{-1}\,\delta/ \chi_{H}^2$ from the density field $\delta$:
\be
 \tau(\bmath\theta)=\frac{3\,\Omega_\mathrm{m}}{c}\, \int _0^{\chi_H} \dd\chi \, a^2\, H(a) \, \frac d{da}\,\frac{D_+}a \, \phi \,(\bmath\theta\chi,\chi)
 \hspace{0.1 cm}.
\ee
Heuristically, the effect originates from an unbalance between the photon's blue-shift when entering a time varying potential well and the red-shift 
experienced at the exit. 

The effect vanishes identically in matter dominated universes $\Omega_\mathrm{m}=1$, since then $D_+/a$ is a constant. Therefore, a non-zero 
iSW-signal will be an indicator of a cosmological fluid with $w\neq0$. After the radiation dominated era it will thus be a valuable tool for investigating 
dark energy cosmologies.

Since the inverse Laplacian which solves for the potential in the Poisson equation introduces a $k^{-2}$ term, small scale fluctuations will be 
quadratically damped. For this reason the iSW-effect provides a signal on large scales and will be negligible above $\ell \approx 100$.

In order to identify the sources of the effect it is sensible to investigate the cross correlation  
of the iSW amplitude with the line of sight projected galaxy density $\gamma$:
\be
 \gamma(\bmath \theta) = b \, \int_0^{\chi_H}\dd\chi\, n(z)\, \frac{dz}{\dd\chi}\, D_+\, \delta\,(\bmath\theta\,\chi , \chi)\hspace{0.1 cm}.
\ee
We obtain the dimensionless observables $\gamma$ and $\tau$ from a line of sight integration of the two dimensionless source fields $\delta$ and 
$\phi$ weighted by functions which carry units of inverse length.

If one is interested in rather small scales one can approximate the sphere locally as beeing plane and perform a Fourier transform
\be
 \gamma(\bmath\ell) = \int \mathrm d ^2 \theta \, \gamma(\bmath\theta)\, \e ^{\displaystyle{- i \,(\bmath{\ell\cdot\theta})}}\hspace{0.1 cm}.
\ee 
Clearly, there is no directional dependence, $\gamma(\bmath \ell)=\gamma(\ell)$, and one can define the spectrum $C_{\gamma\gamma}(\ell)$:
\be
 \bra \gamma(\ell)\, \gamma^{\,*}(\ell^\prime)\ket = (2\pi)^2 \delta_\mathrm{D}(\ell -\ell^\prime) \, C_{\gamma\gamma}(\ell)
\ee
The observable $\tau$ can be transformed in analogous way. With the two weighting functions 
\begin{eqnarray}
 W_{\gamma}(\chi) &=& n(z)\,\frac{H(z)}{c}\, D_+(z) \nonumber  \\
 W_\tau(\chi) &=& 3\,\Omega_{\mathrm m}\, a^2\, \frac Hc \,\frac{\mathrm d}{\mathrm d a}\,\frac{D_+}{a}
\end{eqnarray}
we can now derive the spectra \citep{Limber1953},
\begin{eqnarray}
 C_{\gamma\gamma}(\ell)&=& \int_0^{\chi_H} \frac {\mathrm d \chi}{ \chi ^2}\, W_\gamma ^2(\chi) \, P_{\delta\delta}(k=\ell/\chi)\nonumber \\
 C_{\tau\gamma}(\ell)&=& \int_0^{\chi_H} \frac {\mathrm d \chi}{ \chi ^2}\,W_\tau(\chi)\, W_\gamma(\chi) \, P_{\delta\phi}(k=\ell/\chi)\nonumber \\
 C_{\tau\tau}(\ell)&=& \int_0^{\chi_H} \frac {\mathrm d \chi}{ \chi ^2}\, W_\tau ^2(\chi) \, P_{\phi\phi}(k=\ell/\chi)\hspace{0.1 cm}.
\end{eqnarray}
The power spectra can be related to the density power spectrum:
\be
 P_{\phi\phi}(k) = \frac {P _{\delta\delta}(k)}{(\chi_H\, k)^4} \hspace{0.1 cm}, \hspace{1 cm} 
 P_{\delta\phi}(k) = \frac {P _{\delta\delta}(k)}{(\chi_H\, k)^2}\hspace{0.1 cm}.
\ee
The multiplication factors $k^{-2}$ and $k^{-4}$ tilt the spectra to smaller values for increasing mutipole order $\ell$ and show once again the 
iSW-effect to be a large scale phenomenon.

\section{Tomography with Orthogonal Polynomials}\label{sect_polynomials}

\subsection{Motivation}\label{sect_motivation}
Measurements of the iSW-effect provides integrated information about the structure formation history of our universe since the last scattering surface. 
Cross-correlation with the galaxy density field increases the signal to noise ratio significantly and the spectrum is noiseless 
due to uncorrelated noise in the CMB and the density field.
However, due to the fact that both the cross-correlation spectrum and the galaxy spectrum are line of sight integrated quantities, 
non-linear effects of parameters on the signals could be averaged out and valuable tomographical information would be lost.

Tomographical methods split up the signal from different distances and are therefore able to increase the signal to noise ratio and the sensitvity with 
respect to cosmological parameters. In case of the galaxy spectra this implies that additional covariances between the different spectra have to
be taken into account. 

For a direct tomograpphy in the line of sight integration of the iSW signal the knowledge of the large scale structure potential would be necessary. 
A reconstruction of the potential from the galaxy, however, would not reach the required accuracy.

To circumvent this issue we perform tomography in the galaxy signal and cross-correlate these with the iSW signal. In the course of this we are able 
to withdraw tomographical information also from the iSW signal. We use specifically designed polynomials for a distance weighting 
of the galaxy distribution. Defining the weighted galaxy covariances as a scalar product of the 
polynomials will lead to statistically independent galaxy spectra once the polynomials are orthogonalised. 
This nonlocal binning of the galaxies leads to a diagonalisation of the galaxy signal covariance matrix. The polynomials can then 
also be used for tomographical measurements in the iSW-galaxy cross-correlations.

\begin{figure}
\resizebox{\hsize}{!}{\includegraphics{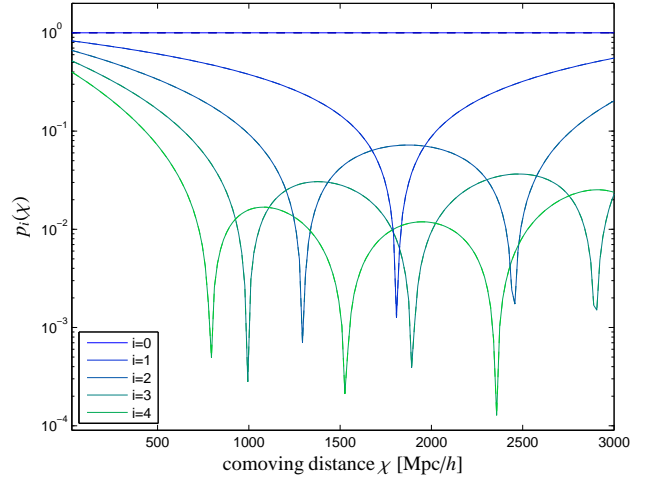}}
\caption{Orthogonal polynomials $p_i(\chi)$, $i=0...4$, as a function of comoving distance $\chi$. The lowest order polynomial is shown in blue, 
the highest order in green. The construction was performed with the Gram-Schmidt algorithm at multipole order $\ell=100$.}
\label{fig_polynomials}
\end{figure}

\subsection{Construction of orthogonal sets of polynomials}\label{sect_construction}
Weighting the given galaxy distribution function $n(\chi)=n(z)\,\dd z/\dd\chi=n(z)\,H(z)/c$ with a polynomial $p_i(\chi)$ modifies the galaxy weighting 
function to
\begin{equation}
 W_{\gamma}^{(i)}(\chi) = p_i(\chi)\,W_{\gamma}(\chi) = p_i(\chi)\,n(z)\,\frac{H(z)}{c}\, D_+(z) \hspace{0.1 cm}.
\end{equation}
For the polynomials $p_i(\chi)$ and $p_j(\chi)$ we require orthogonality
\begin{equation}
 \bra p_i, p_j \ket = 0  \hspace{0.1 cm} \hbox{for} \hspace{0.1 cm}  (i \neq j)
\end{equation}
with respect to the following scalar product for the polynomials:
\begin{equation}
 \bra p_i, p_j \ket \equiv S_{\gamma\gamma}^{(ij)}(\ell) \equiv \int_0^{\chi_H}\, \frac{\mathrm \dd\chi}{\chi^2}\, 
W_{\gamma}^{(i)}(\chi)\, W_{\gamma}^{(j)}(\chi)\, P(k=\ell/\chi)\hspace{0.1 cm}.
\end{equation}
The necessary properties for a scalar product are obviously fullfilled
($\bra p_i, p_i \ket \geq 0$,  $\bra p_i, p_i \ket = 0\,\Leftrightarrow\, p_i \equiv 0$ and linearity).
We use the Gram-Schmidt procedure to construct orthogonal polynomials out of the family of monomials 
\begin{equation}
 p_i^\prime(\chi)=\left(\frac{\chi}{\chi_{\mathrm{node}}}\right)^i\hspace{0.1 cm},
\end{equation} 
where $\chi_{\mathrm{node}}$ sets the position of the node of the first polynomial, which is in our case set to the median value of the redshift 
distribution. However, a change in $\chi_\mathrm{node}$ is completely absorbed in the coefficient and has no influence on the polynomials. 
Starting with the zero-order polynomial
\begin{equation}
p_0(\chi) = p_0^{\prime}(\chi) \equiv 1 \hspace{0.1 cm},
\end{equation}
the polynomials are constructed iteratively,
\begin{equation}
 p_i(\chi) = p_i^{\prime}(\chi) - \sum_{j=0}^{i-1}\frac{\bra p_i^{\prime},p_j\ket}{\bra p_j, p_j\ket}\, p_j(\chi)\hspace{0.1 cm}.
\end{equation}
The procedure has to be performed for every multipole $\ell$. The index $\ell$ of the polynomials $p_i(\chi)$ has been omitted for clarity. 
As one can see, the zero-order  scalar product is equal to the galaxy spectrum:
\begin{equation}
 \bra p_0, p_0 \ket = S_{\gamma\gamma}(\ell) \hspace{0.1 cm}.
\end{equation}
Therefore, the unweighted case is already contained in the first weighting function. 
\begin{figure}
\resizebox{\hsize}{!}{\includegraphics{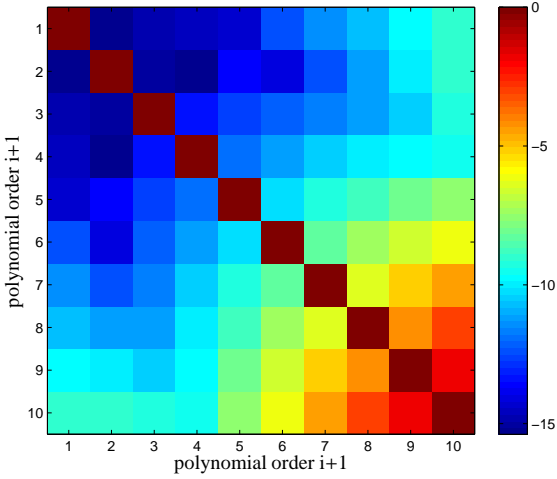}}
\caption{Numerical accuracy for the orthogonality relation $\bra p_i,p_j\ket$ at $\ell=20$ in logarithmic representation. The accuray imposes a limit 
on the number of included polynomials.}
\label{fig_ortho}
\end{figure}
Finally, we can weight also the tracer density modes $\gamma(\ell)$ themselves with a polynomial $p_i(\chi)$
\be
 \gamma^{(i)}(\ell)= \int _0^{\chi_H} \dd\chi\, W_\gamma ^{(i)}(\chi)\, \delta
\ee
for which a generalized version of the well known expression for the covariance holds in case of homogeneous and isotropic random fields:
\be
 \bra \gamma^{(i)}(\ell) \, \gamma^{(j)*} (\ell^\prime)\ket = (2\pi)^2 \, \delta_D (\ell-\ell^\prime)\, S_{\gamma\gamma}^{(ij)}(\ell) \hspace{0.1 cm}
\ee
with $S_{\gamma\gamma}^{(ij)}(\ell)\propto \delta_{ij}$.

\subsection{Properties of orthogonal polynomials}\label{sect_properties}
In Fig.~\ref{fig_polynomials} the orthogonal polynomials are shown up to a polynomial order of $i=4$. They show an increasing number of zero points 
roughly at the positions where the previous polynomial reaches a local maximum or minimum, which intuitively indicates their orthogonality. 

As one can see in Fig.~\ref{fig_ortho} orthogonality is fulfilled until numerical limitations become significant at a polynomial order of $q\approx9$. 
The inreasing numerical deviations from the orthogonality condition ($\bra p_i,p_j\ket = 0$ for $ i\neq j$) is due to the iterative method, 
which cumulates errors throughout the process. This implies the accuraccy to shrink from $10^{-15}$ for $i=0$ to $10^{-3}$ for $i=8$. 
This is a well known disadvantage of the Gram-Schmidt orthogonalisation method, especcially when dealing with functions as opposed to vectors, since 
there ist larger numerical noise in the evaluation of the scalar products.
However, as we will later see, it is not necessary for our application to go to even higher orders.

\begin{figure}

\resizebox{\hsize}{!}{\includegraphics{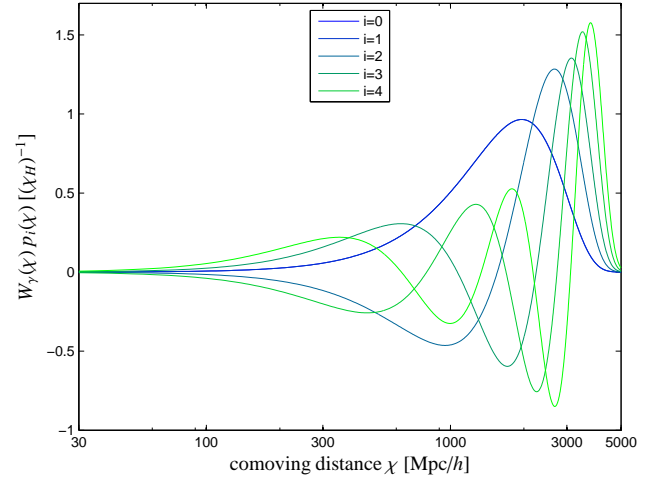}}
\caption{Weighted galaxy efficiency function $W_\gamma^{(i)}(\chi)$, $i=0...4$, as a function of comoving distance at multipole order $\ell=20$.}
\label{fig_weighting_g}
\end{figure}

In Fig.~\ref{fig_weighting_g} the weighted galaxy efficiency functions $W_{\gamma}^{(i)}(\chi)$ are depicted, which are modified by the polynomials
$p_i(\chi)$ at an angular scale of $\ell=19$. The case $i=0$ refers to the weighting function without tomography, $W_{\gamma}^{(0)}(\chi)= W_\gamma(\chi)$.
One can easily observe the order of the polynomial hierarchy at the high distance end of the functions, where one after another approaches zero.

%

\begin{figure}
\resizebox{\hsize}{!}{\includegraphics{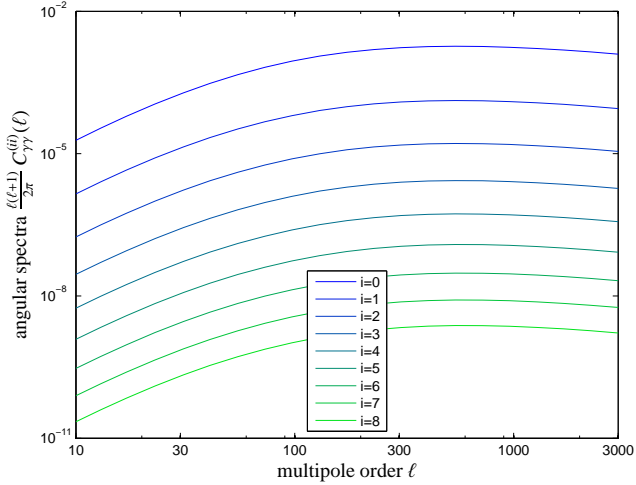}}
\caption{Pure galaxy-galaxy spectra $S_{\gamma\gamma}^{(ii)}(\ell)$, $i=0...8$, weighted with orthogonal polynomials $p_i(\chi)$, 
as a function of the multipole order $\ell$. 
$S_{\gamma\gamma}^{(00)}(\ell)$ (blue) refers to the non-tomographic spectrum $S_{\gamma\gamma}(\ell)$. 
One can see a decrease in amplitude for increasing multipole order $\ell$.} 
\label{fig_spectra_gg}
\end{figure}

\begin{figure}
\resizebox{\hsize}{!}{\includegraphics{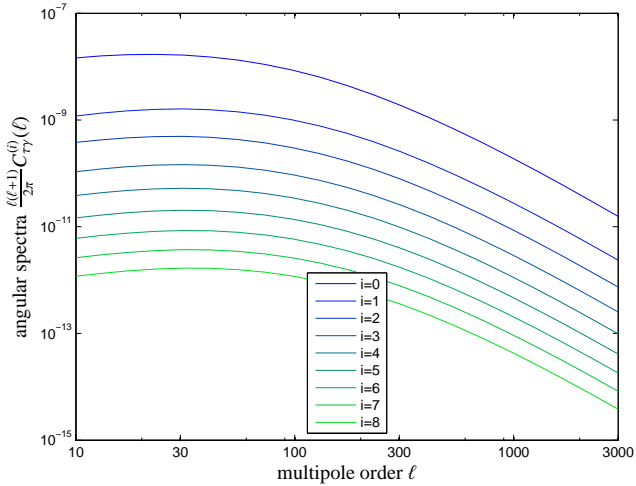}}
\caption{Galaxy-iSW cross-spectra $C_{\tau\gamma}^{(i)}(\ell)$, $i=0...8$, weighted with orthogonal polynomials $p_i(\chi)$, 
as a function of the multipole order $l$. $C_{\tau\gamma}^{(0)}(\ell)$ (blue) refers to the non-tomographic spectrum $C_{\tau\gamma}(\ell)$.}
\label{fig_spectra_tg}
\end{figure}

The modified spectra $C_{\gamma\gamma}^{(ii)}(\ell)$ and $C_{\tau\gamma}^{(i)}(\ell)$ are shown in Fig.~\ref{fig_spectra_gg} 
and Fig.~\ref{fig_spectra_tg}, respectively. 
The drop in amplitude is mainly an effect of the absence of normalisation, while one can in fact observe slight differences in shape. 
However, these differences are small, since the polynomials only mildly depend on the multipole order $\ell$. Therefore, the overall shape of the 
spectra is still dominated by the zero-order spectra $C_{\gamma\gamma}^{(00)}(\ell)$ and $C_{\tau\gamma}^{(0)}(\ell)$, respectively. Thanks to 
the orthogonalisation these spectra now provide statistically independent information. In the next section we aim to combine singals from the 
galaxy distribution, the iSW-effect and the cross-spectra to investigate statistical bounds on cosmological parameters.

\section{Statistics}\label{sect_statistics}
This section aims to connect cosmic variance and statistical noise with the iSW-signal and its cross-correlations into a meaningful 
statistical formulation. In the course of this we construct covariance matrices for the polynomial-weighted spectra. Statistical errors on 
cosmological parameters are estimated in the Fisher-matrix formalism. 
Furthermore, we investigate the signal strength of the different spectra and their dependence on the number of polynomials used.

\subsection{Variances of galaxy number counts}\label{sect_covariance}
For forecasting statistical errors, we need to derive expressions for the signal covariance and noise.
We will start from a discrete formulation with a set of weighting coefficients $w_m$ for the counted galaxy number $m$. Clearly, the weighting coefficient 
$w_m$ will depend on the distance of the respective galaxy. Later, we will generalise the 
formalism to the continuous case, in which the weighting procedure is performed by the polynomials $p_i(\chi)$.
The standard deviation $\sigma_{ww}$ of a weighted galaxy count with weighting coefficients $w_m$ is given by
\be
 \sigma^2_{ww}=\frac{1}{\sum_m w_m \, \sum_n w_n} \,\sum_{m,n} w_m\,w_n\, \delta_{mn} \hspace{0.5 cm}
\ee
which reduces to a Poissonian result in the case of $w_m$ being either $0$ or $1$:
\be
 \sigma^2_{ww}=\frac{1}{\overline{n}} \hspace{0.5cm} \hbox{with} \hspace{0.5cm} \overline{n} = \sum _n w_n \hspace{0.1 cm}.
\ee
The counted quantity $\overline{n}$ in our case is defined as the mean density of galaxies per steradian, for which we will substitute 
$\overline{n}=40/\mathrm{arcmin}^2$, which is characteristical for the EUCLID galaxy survey. Considering two different sets of weighting 
factors $w_m$ and $v_n$, we generalise the standard deviation to
\be
 \sigma^2_{wv}=\frac{1}{\sum_m w_m \, \sum_n v_n} \,\sum_{m} w_m\,v_m \hspace{0.1 cm},
\ee
which will in the continuum limit be a cross variance weighted with two different polynomials. For the continuum limit the transition
\be
 \sum_m \hbox{...} \rightarrow \overline{n}\, \int d \,\chi n(\chi) \hbox{...} 
\ee
is performed which conserves the total number count $\overline{n}$ due to the unit normalised galaxy distribution function $n(\chi)$.
The discrete weighting sets $w_m$ and $v_n$ are then represented by $p_i(\chi)$ and $p_j(\chi)$ so that the noise covariance in the continuous case reads
\be
 N_{\gamma\gamma}^{\,(ij)} (\ell) \equiv \sigma_{ij}^2 =\frac {1}{\overline{n}} \,
\frac{\int \dd\chi \, n(\chi) \, p_i(\chi)\, p_j(\chi)}{\int \dd\chi\, n(\chi)\,p_i(\chi)\,\, 
  \int \dd\chi \,n(\chi) \, p_j(\chi)} \hspace{0.1 cm}.
\label{noise}
\ee
The noise term $N_{\gamma\gamma}^{\,(ij)} (\ell)$ still depends on $\ell$ since the polynomials are constructed for each multipole order separately.
We omit the $\ell$-dependence of the polynomials $p_i(\chi)$ for clarity.
Eqn.~(\ref{noise}) motivates the following choice of normalisation for our polynomials:
\be
 p_i(\chi) \leftarrow \frac{p_i(\chi)}{\int \dd\chi\,  n (\chi) \, p_i(\chi)}\hspace{0.1 cm} .
\ee
In this normalisation the galaxy number noise reads
\be
  N_{\gamma\gamma}^{\,(ij)} (\ell) =\frac {1}{\overline{n}} \,\int \dd\chi \, n(\chi) \, p_i(\chi)\, p_j(\chi) \hspace{0.1 cm} ,
\ee
while the galaxy spectrum can be written as
\be
 S^{(ij)}_{\gamma\gamma} (\ell)= \int_0 ^{\chi_H} \frac{\dd\chi}{\chi^2}\, W_\gamma^{(i)}(\chi)\, W_\gamma^{(j)}(\chi) \,P (k=\ell/\chi)
\ee
The limitation in polynomial order due to increasing noise in the polynomials $p_i(\chi)$ can already be illustrated: 
Since $n(\chi)$ is a slowly varying function the rapid oscillations of high order polynomials will drive the values of 
the integrals $\int \dd\chi\,  n (\chi) \, p_i(\chi)$ to smaller numbers and will therefore increase the noise in $p_i(\chi)$.
We point out that for order zero the non-tomographic case is recovered, giving the standard Poissonian expression for the noise 
$N_{00}=1/\overline{n}$ and the integrated galaxy spectrum in the signal part $S^{(00)}_{\gamma\gamma}(\ell) =S_{\gamma\gamma}(\ell)$.

While the orthogonalisation procedure leads to a diagonal galaxy signal covariance, the noise part will not be diagonal any more: 
$N_{\gamma\gamma}^{(ij)} \neq 0$ for $i\neq j$. In contrary to this method, a traditional binning in $z$ would lead to a diagonal noise contribution 
and off-diagonals in the signal part.
\subsection{Fisher analysis}\label{sect_fisher}
In order to use both iSW signals and galaxy spectra in our Fisher analysis, we now define an extended data vector
\be
 \bmath{x}\,(\ell)=  \left(\begin{tabular}{r}$\tau\,\,(\ell)$\\
		    $\gamma^{(i)}(\ell)$
		      \end{tabular}\right)  \hspace{0.1 cm}.
\ee
The total covariance matrix, $C(\ell) = S(\ell) + N(\ell)$, for these data vectors is block-diagonal due to the independence of the $\ell$-modes: 
Each block 
\be
 C(\ell) = 
     \left(\begin{tabular}{c | c}$C_{\tau\tau}(\ell)$&$C_{\tau\gamma}^{\,(j)}(\ell)$ \\  $C_{\tau\gamma}^{\,(i)}(\ell)$&$C_{\gamma\gamma}^{\,(ij)}(\ell)$
     \end{tabular}\right)  
\ee
consists of a signal part 
\be
 S(\ell) = 
     \left(\begin{tabular}{c | c}$S_{\tau\tau}(\ell)$&$C_{\tau\gamma}^{\,(j)}(\ell)$ \\  $C_{\tau\gamma}^{\,(i)}(\ell)$&$S_{\gamma\gamma}^{\,(ij)}(\ell)$
     \end{tabular}\right)  \hspace{0.1 cm },
\ee
where $S_{\gamma\gamma}^{(ij)}(\ell)\propto \delta_{ij}$ by construction,
and a noise contribution
\be
  N(\ell) = 
     \left(\begin{tabular}{c | c}$N_{\tau\tau}(\ell)$&$0$\\  $0$&$N_{\gamma\gamma}^{\,(ij)}(\ell)$
     \end{tabular}\right) \hspace{0.1 cm}.
\ee
with polynomial orders $0 \leq i,j \leq q$.
Due to uncorrelated noise in the CMB and the galaxy density field the noise of the cross-spectra vanishes. 
The CMB part consists of the pure iSW signal 
\be
 S_{\tau\tau}(\ell) = \int_0^{\chi_H} \frac{\dd\chi}{\chi^2} \, W_\tau ^2 (\chi) \, P_{\phi\phi}(k = \ell/\chi)
\ee
with $P_{\phi\phi}(k)= P_{\delta\delta}/(\chi_H\, k)^4$ while the noise can be split into the primary CMB fluctuations $C_{\mathrm{CMB}}(\ell)$ 
and an instrumental noise term $C_{\mathrm{beam}}(\ell)$
of PLANCK \citep{PlanckCollaboration2011}:
\be
 N_{\tau\tau}(\ell) = C_{\mathrm{CMB}}(\ell) + w^{-1}\,\exp{\left(-\Delta\theta^2\, \ell^2\right)}\hspace{0.1cm},
\ee
with the beam size $\Delta \theta = 8.77\, \times 10^{-4}$ and the squared pixel noise $w^{-1}=0.2\, \mu \mathrm K / T_\mathrm{CMB}$ \citep{Knox1995}. The noiseless cross spectra are formed by one modified weighting 
function only:
\be
 C_{\tau\gamma}^{(i)}(\ell) = \int_0^{\chi_H} \frac{\dd\chi}{\chi^2} \, W_\tau (\chi)\, W_\gamma^{(i)} (\chi)\,P_{\delta\phi}(k = \ell/\chi) 
\ee
with $P_{\delta\phi}(k)= P_{\delta\delta}/(k\, \chi_H)^2$. We point out that only the galaxy part of the signal covariance was diagonalised by our method. 
Consequently, the cross-spectra $C_{\tau\gamma}^{(i)}(\ell)$ are the only off-diagonal entries in the covariance matrix.
 
The likelihood for observing these Gaussian-distributed modes $\bmath{x}\,(\ell)$ for a given parameter set $\bmath p$ is defined as (Tegmark, 1997):
\be
 \mathcal{L}(\bmath x (\ell)\,|\,\bmath p) = \frac {1}{\sqrt{(2\pi)^N\,\det{C(\ell)}}}
  \, \exp{\left( -\frac 12  \bmath x ^{\,T}(\ell) \,\, C^{-1} (\ell)\,\, \bmath x^{\,*}(\ell) \right)}\hspace{0.1cm}.
\ee

\begin{figure}
\resizebox{\hsize}{!}{\includegraphics{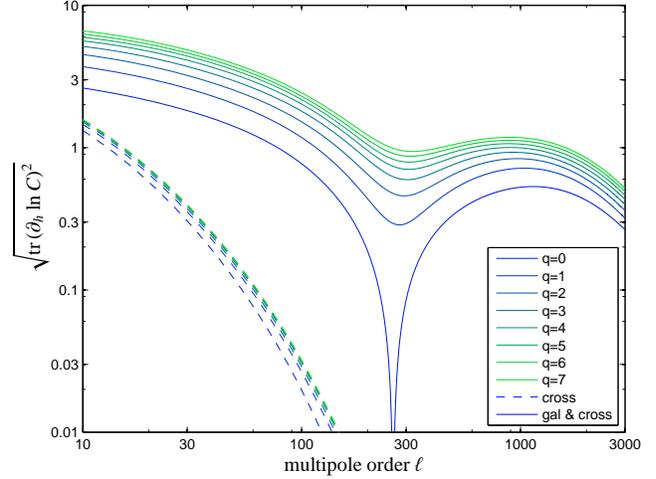}}
\caption{Sensitivity $\sqrt{\trace \left(\partial_{h} \ln C(\ell)\right)^2}$ of the Fisher matrix with respect to the Hubble parameter 
$h$ as a function of the multipole order $\ell$, and cumulative polynomial order $q$. Sensitivities are shown with derivatives of all 
spectra taken into account (solid lines) in comparison to the case where only the cross-spectra were considered in the signal part 
(dashed lines). For the covariance the survey properties of EUCLID have been assumed.}
\label{fig_derivative_h}
\end{figure}

\begin{figure}
\resizebox{\hsize}{!}{\includegraphics{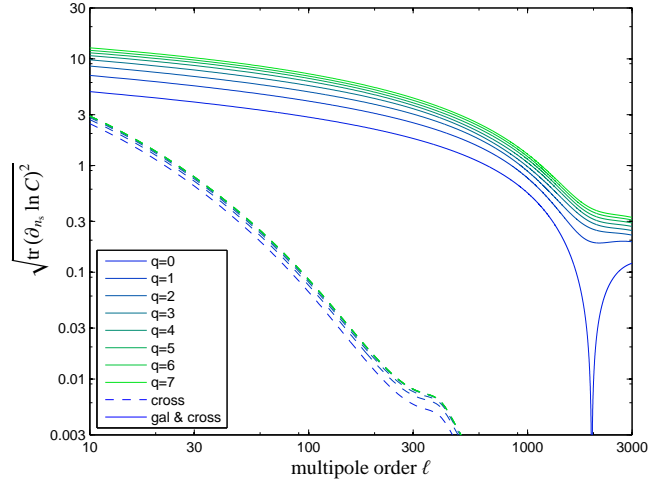}}
\caption{Sensitivity $\sqrt{\trace \left(\partial_{n_{\mathrm s}} \ln C(\ell)\right)^2}$ of the Fisher matrix with respect to the initial 
slope of the power spectrum $n_{\mathrm s}$ as a function of the multipole order $\ell$, and cumulative polynomial order $q$. Sensitivities are shown with derivatives of all 
spectra taken into account (solid lines) in comparison to the case where only the cross-spectra were considered in the signal part 
(dashed lines). For the covariance the survey properties of EUCLID have been assumed.}
\label{fig_derivative_ns}
\end{figure}

\begin{figure}
\resizebox{\hsize}{!}{\includegraphics{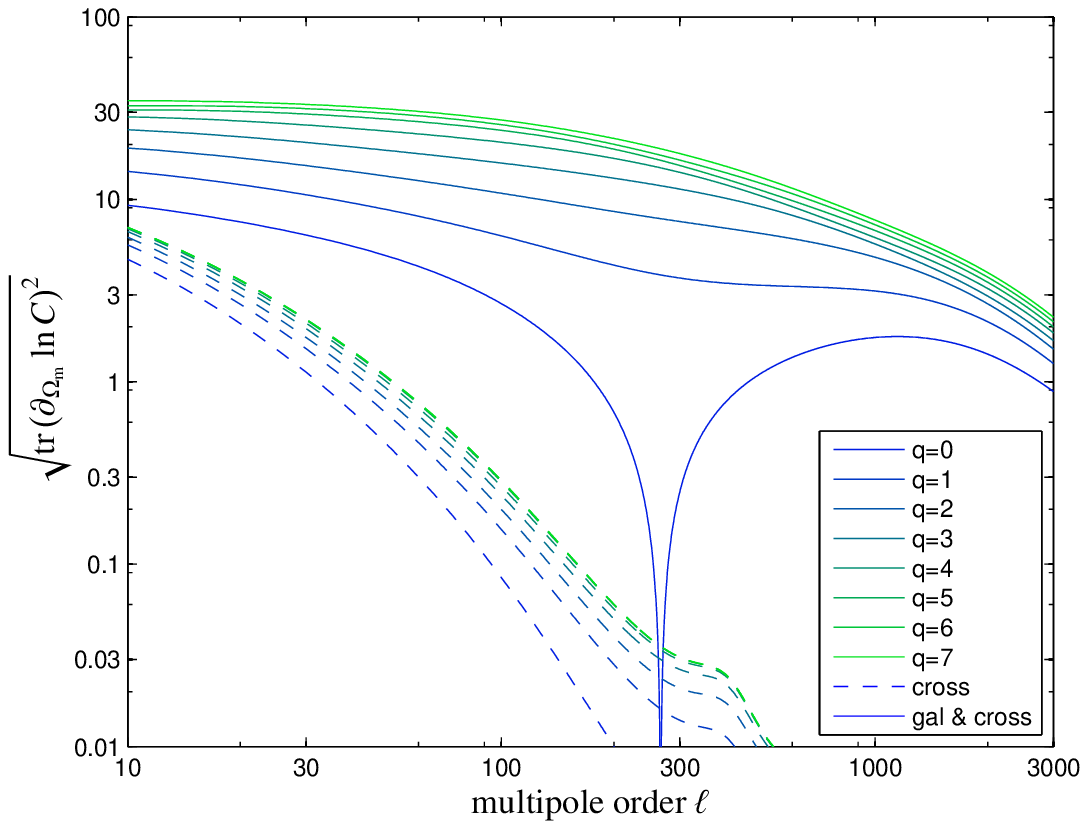}}
\caption{Sensitivity $\sqrt{\trace \left(\partial_{\,\Omega_{\mathrm{m}}} \ln C(\ell)\right)^2}$ of the Fisher matrix with respect to the 
matter density parameter $\Omega_{\mathrm{m}}$ as a function of the multipole order $\ell$, and cumulative polynomial order $q$. Sensitivities are shown with derivatives of all 
spectra taken into account (solid lines) in comparison to the case where only the cross-spectra were considered in the signal part 
(dashed lines). For the covariance the 
survey properties of EUCLID have been assumed.}
\label{fig_derivative_omegam}
\end{figure}

Defining the data matrix as $D_{ij}(\ell) = x_i(\ell)\,x_j(\ell)$ with $\bra D\ket= C$ and using the relation, $\ln\det {(C)} =\trace \ln {(C)}  $ 
one can write the $\chi^2$-functional $\mathcal L\,\propto \exp\,(-\chi^2/2)$, with help of the logarithmic likelihood $L \equiv -\ln \mathcal L$:
\be
 \chi^2 = -2\, L = \trace\sum_\ell \left[\ln C + C^{-1}\, D \right] \hspace{0.1 cm}.
\ee 
Each multipole $\ell$ provides $(2\ell+1)$ independent $m$-modes. If we interprete $\mathcal L$ as a Bayesian probability, 
the local behaviour of the likelihood function around the point of maximum likelihood is 
determined by the Hesse matrix of $L$ at this point:
\be
 (C^{-1})_{\mu\nu} \equiv \frac {\partial^2 L}{\partial p_\mu \,\partial p_\nu}
\ee
The Fisher information matrix is then given as the expectation value of this quantity summed over all multipole orders $\ell$:
\be
 F_{\mu\nu} = \left\bra \frac{\partial^2 L}{\partial p_\mu\, \partial p_\nu} \right\ket = \sum_\ell \frac{2\ell+1}{2}
    \trace\left( \frac {\partial}{\partial p_\mu}\, \ln C(\ell)\, \frac {\partial}{\partial p_\nu}\, \ln C(\ell) \right)\hspace{0.1 cm }.
\ee
For each $\ell$ the $(2\,\ell+1)/2$ $m$-modes provide statistically independent information.
In the course of our Fisher matrix calculations we will work in the limit $\partial S_{ij}/\partial p_\mu \gg \partial N_{ij}/\partial p_\mu$
and therefore neglect the noise dependence on the cosmological parameters. This approximation is well justified in our case.

Next we have a look at the ratio of the sensitivities of the spectra with respect to cosmological parameters and the covariance. This quantity 
equals the contribution of a certain multipole $\ell$ to the respective Fisher matrix diagonal element:
\be
 \sqrt{\trace \left(\frac{\partial \ln C(\ell)}{\partial \p_\mu}\right)^2} = \sqrt{\frac{2}{2\ell+1}}\frac {\mathrm d F_{\mu\mu}}{\mathrm d \ell}
\hspace{0.1 cm }.
\ee
In Fig.~\ref{fig_derivative_h} - Fig.~\ref{fig_derivative_omegam} we show these sensitivities in solid lines for the full information from galaxy spectra, 
cross-spectra and iSW-effect included for the parameters $h$, $n_{\mathrm s}$ and $\Omega_{\mathrm m}$, respectively.
At zero order they all exhibit a certain $\ell$-range at which the covariance is insensitive to variations of the respective cosmological parameter. 
Naturally, angular scales in the vicinity of this zero point do not contribute much sensitivity to the Fisher matrix.
This effect is cured if we include all polynomials $0\leq i \leq q$. The combination of multiple line of sight-weighted measurements 
lifts the sensitivities at these points continuously with increasing number of involved polynomials until the effect saturates.

For multipole orders $\ell$ reaching higher values ($\ell \approx 3000$) the sensitivity starts to drop rapidly. On these small scales the noise contribution 
begins to dominate and the Fisher matrix ceases to grow further. 

In dashed lines the sensitivities are shown if only the cross-spectra are included 
in the derivatives. Again the sensitivities grow with increasing cumulative polynomial order $q$, although in this case the zero order sensitivity does not suffer from 
any singular effects. Characteristic properties of the iSW-effect are recovered showing it to be a large scale effect due to the $k^{-2}$ proportionality
originating in the Poisson equation. Above moltipoles of about $l\approx 100$ the information provided by the cross spectra becomes negligible. 
Clearly, the cross-spectra Fisher matrix is most sensitive to the matter density parameter $\Omega_{\mathrm m}$,
which shows the strongest increase in sensitivity for increasing cumulative polynomial order $q$.

\subsection{Signal to noise ratio}\label{sect_s2n}
\begin{figure}
\resizebox{\hsize}{!}{\includegraphics{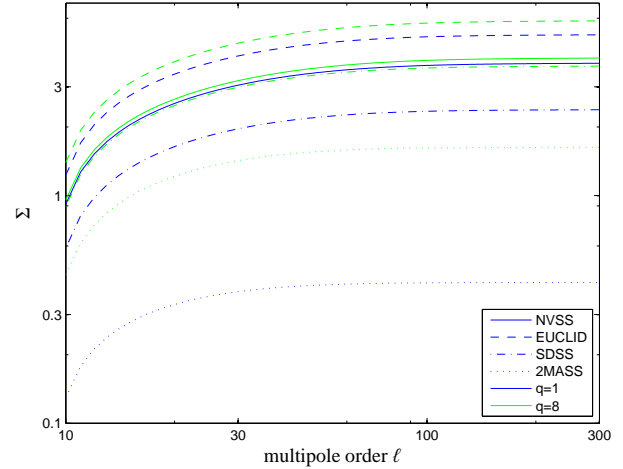}}
\caption{Cumulative signal to noise ratio $\Sigma$ depending on the multiple-order $\ell$ for the survey characteristics of 
2MASS (dotted), SDSS (dashed-dotted), EUCLID (dashed) and NVSS (solid). Shown is the improvement between cumulative polynomial order $q=1$ (blue) and $q=8$ (green).}
\label{fig_s2n}
\end{figure}

A signal's power to constrain cosmological parameters is most reliably quantified by the signal to noise ratio
\be
 \Sigma^2 = f_{\mathrm{sky}}\sum_{\ell} \, \frac{2\, \ell +1 }{2}\trace{\left( C^{-1}(\ell)\, S(\ell) \right)^2}\hspace{0.1 cm }.
\ee
Besides the physical process the signal to noise ratio strongly depends on the characteristics of the survey at hand. In the case of galaxy surveys the 
most important survey parameters are the sky coverage $ f_{\mathrm{sky}}$ and the median redshift $ z_{\mathrm{med}}$. In Fig~\ref{fig_s2n} the signal to noise 
ratios for the survey characteristics of EUCLID, 2MASS \citep[$z_\mathrm{med}\approx 0.1$,][]{Afshordi2004}, SDSS \citep[$z_\mathrm{med}\approx 0.5$,][]
{Bielby2010} and NVSS \citep[$z_\mathrm{med}\approx 1.2$,][]{Boughn2005} are shown. Clearly, also the 
signal to noise ratio increases for higher polynomial orders due to the diagonal structure of the signal covariance.
We find an improvement of $\approx 16 \%$ in the signal to noise ratio between cumulative polynomial order $q=1$ and $q=8$. 
As expected, at a multipole order of a few hundred the cumulative signal strength saturates as a result of the Poissonian $k^{-2}$ damping term in the iSW-effect.
The actual realisation of the matter distribution in the observed universe introduces a systematic noise in the iSW detections 
known as local variance. Using the so called optimal method one can decrease this bias by working conditional on the large scale structure data 
and gain 7\% in signal to noise ratio \citep{Frommert2008}. The tomographical method presented in our work should be also 
applicable to the reconstructed large scale structure used in the optimal method. Therefore, a combination of these two methods would be sensible.

\begin{figure}
\resizebox{\hsize}{!}{\includegraphics{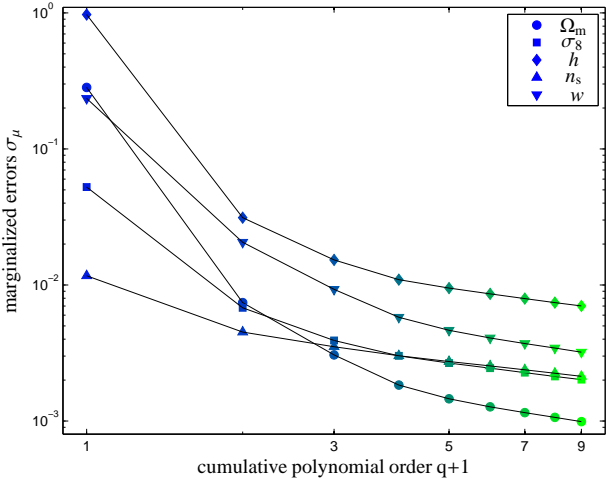}}
\caption{Lower limits on the marginalized statistical errors $\sigma_\mu$ on the estimates of the cosmological parameters 
$\Omega_{\mathrm{m}}$ (circles), $\sigma_8$ (squares), $h$ (lozenges), $n_{\mathrm s}$ (triangles, pointing up) and $w$ (triangles, pointing down) derived 
from the Cram\'er-Rao inequality, as a function of the cumulative polynomial order $q$. 
The Fisher matrix was derived including the derivatives of all spectra 
$C_{\gamma\gamma}^{(ii)}(\ell)$, $C_{\tau\gamma}^{(i)}(\ell)$ and $C_{\tau\tau}(\ell)$. 
Again, the EUCLID survey characteristics have been used.}
\label{fig_error_marg}
\end{figure}

\subsection{Statistical errors}\label{sect_errors}
The Cram\'{e}r-Rao inequality introduces a lower limit on the marginalized standard deviation of the estimated cosmological parameters. 
These are given by the diagonal elements of the inverse Fisher matrix:
\be
 \sigma_\mu \geq \left( (F^{-1})_{\mu\mu}\right)^{\frac 12} \hspace{0.1 cm}.
\ee
In Fig.~\ref{fig_error_marg} these errros are depicted for the five cosmological parameters $\Omega_{\mathrm{m}}$, $\sigma_8$, $h$, $n_{\mathrm s}$ and $w$. The plot follows 
the evolution of the errors with increasing number of included polynomials $q$. While for small polynomial orders the Cram\'{e}r-Rao errors decrease 
rapidly with different characteristics for each parameter, the improvement slows down for higher order polynomials and assumes a characteristical behaviour.
This behaviour can approximately be described by the inverse root of the polynomial order $\sigma_\mu\propto 1/\sqrt q$. A similar characteristic was found 
in the application of this method to the weak lensing shear spectra \citep{Schaefer2011}. Clearly, the cosmological parameter $\Omega_{\mathrm{m}}$ profits 
most of the tomographic method, which was already indicated by its sensitivity improvement discussed in Section~\ref{sect_fisher}. Going to even higher orders 
is difficult due to cumulative errors in the Gram-Schmidt orthogonalisation method.

\begin{figure}
\resizebox{\hsize}{!}{\includegraphics{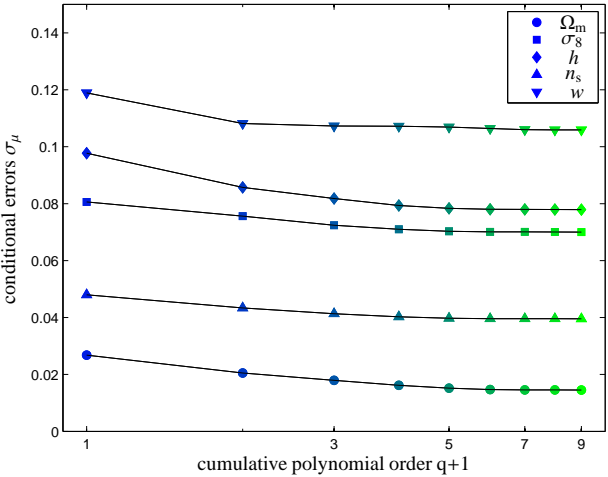}}
\caption{Conditional statistical errors $\sigma_{\mu,\mathrm{con}}$  on the estimates of the cosmological parameters 
$\Omega_{\mathrm{m}}$ (circles), $\sigma_8$ (squares), $h$ (lozenges), $n_{\mathrm s}$ (triangles, pointing up) and $w$ (triangles, pointing down). 
The Fisher matrix was derived including the derivatives of the cross-spectra $C_{\tau\gamma}^{(i)}(\ell)$ only, 
EUCLID survey characteristics have been used.}
\label{fig_error_cond}
\end{figure}

If we are interested in how a single cosmological parameter can be constrained assuming that all other parameters are fixed, we have to study the 
conditional errors. These can be obtained from the inverse diagonal elements of the Fisher matrix
\be
 \sigma_{\mu \mathrm{,con}} = (F_{\mu\mu})^{-\frac 12} \hspace{0.1 cm}.
\ee
For studying the improvement provided by the cross-spectra, we plot the conditional errors as a function of cumulative polynomial order $q$ in 
Fig.~\ref{fig_error_cond}. Here, only the derivatives of the cross-spectra were taken into account in the Fisher matrix calculation. 
Again $\Omega_{\mathrm{m}}$ is subjected to the strongest improvement, its conditional error decreases by $\approx 30\%$.
In contrast to the marginalized errors the evolution of the conditional errors does not show a $1/\sqrt{q}$ behaviour but rather saturates at 
polynomial order of $q\approx 5$.

\begin{figure}
\resizebox{\hsize}{!}{\includegraphics{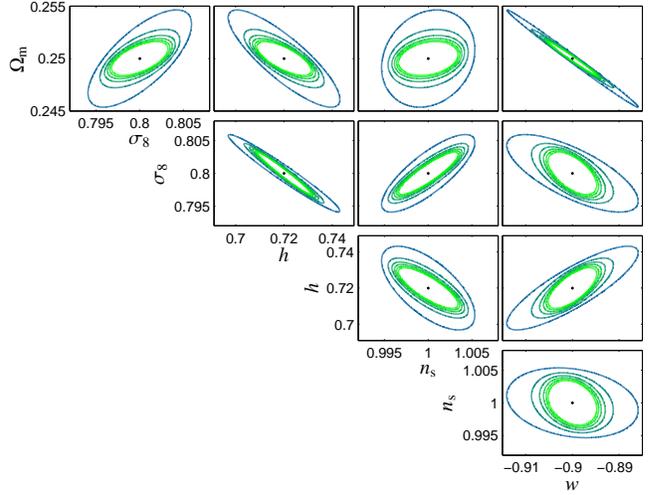}}
\caption{The 2-dimensional $1\sigma$-error ellipses for the cosmological parameters $\Omega_{\mathrm{m}}$, $\sigma_8$, $h$, $n_{\mathrm s}$ and $w$ from 
EUCLID using tomography with orthogonal polynomials are shown in this plot. The $1\sigma$ confidence regions decrease in size with 
increasing number of included polynomials, reaching from $q=2$ (blue) to $q=8$ (green). The ellipses are evaluated with a maximum multipole order of 
$\ell_{\mathrm{max}}=3000$.} 
\label{fig_fisher_ellipse}
\end{figure}

Finally, we are interested in the 2-dimensional marginalized logarithmic likelihood $\chi_{\mathrm m}^2$ around the fiducial model 
$\bmath p _{\mathrm{fid}}$
\be
 \chi^2_{\mathrm m} = \left(\begin{tabular}{c}$p_\mu - p_{\mu\mathrm{,fid}}$\\
			     $p_\nu - p_{\nu\mathrm{,fid}}$
                            \end{tabular}\right)
\left(\begin{tabular}{cc}$(F^{-1})_{\mu\mu}$&$(F^{-1})_{\mu\nu}$\\
			     $(F^{-1})_{\nu\mu}$&$(F^{-1})_{\nu\nu}$
                            \end{tabular}\right)
\left(\begin{tabular}{c}$p_\mu - p_{\mu\mathrm{,fid}}$\\
			     $p_\nu - p_{\nu\mathrm{,fid}}$
                            \end{tabular}\right)
\ee
for which the $1\sigma$-error ellipses are depicted in Fig.~\ref{fig_fisher_ellipse}. Starting with $q=2$, we have combined up to nine polynomials. 
Besides the expected shrinking of the ellipses for higher numbers of included polynomials, it is interesting to see how the degeneracies slightly change 
their orientations in the course of tomographic improvement. This is very likely due to distance dependencies of the signal sensitivities.

\section{Summary}\label{sect_summary}
In this paper a tomographic method for measuring iSW-galaxy cross-spectra and galaxy spectra has been presented. It has been carried out by 
sight-weighting of the iSW-effect and the tracer density field with specifically constructed orthogonal polynomials.

(i) The Gram-Schmidt orthogonalisation procedure has been used to construct orthogonal polynomials in order to diagonalise the weighted galaxy signal 
covariance. The method projects out statistically independent signal contributions at the price of off-diagonals in the noise part.
It differs from traditional tomographical approaches, for instance from most tomographical techniques in weak lensing measurements, 
in which the noise part is diagonalised. 
Due to cumulative numerical errors with increasing polynomial order, this method is limited to order $i\approx8$.    

(ii) The improvement of the signal to noise ratios with cumulative polynomial order was investigated for 
the galaxy surveys 2MASS \citep{Afshordi2004}, SDSS \citep{Bielby2010}, NVSS \citep{Boughn2005} and EUCLID. The signal to noise 
ratio for the cross-spectra only has been improved by 16\% at a cumulative polynomial order of $q=8$.

(ii) A Fisher-matrix analysis was used to forcast how well cosmological parameters can be constrained by different galaxy surveys, 
combining signals from the iSW-effect as well as from the tracer density field. 
The marginalised errors show simple inverse square-root behaviour with increasing number of included polynomials, which can be interpreted also as a sign of the 
statistical independent signal contributions. Conditional errors on parameters contrained only by the cross-spectra decrease by up to $\approx 30\%$ 
in case of the matter density parameter $\Omega_\mathrm{m}$.  

(iv) While for the cross-spectra only the conditional errors show a saturation already at quite low number of included polynomials $q\approx5$, 
it would still be worth improving this method in order to reach even higher orders, since marginalised errors for the full signal did not 
yet saturate at cumulative order of $q=8$.

(v) Using the wrong model in the construction of the polynomials can introduce an estimation bias on cosmological parameters. This effect was thoroughly studied in a similar approach for
weak lensing measurements \citep{Schaefer2011}. In most cases the bias was found to be small compared to the statistical errors. Since, in addition, 
iteration between parameter estimation and polynomial construction is able to further reduce this bias, a wrongly chosen cosmology appears 
unlikely to affect measurements.
\section*{Acknowledgements}
Our work was supported by the German Research Foundation (DFG) within the framework of the Priority Programme 1177 and the excellence 
initiative through the Heidelberg Graduate School of Fundamental Physics. 
We also thank Patricio Vielva and Carlos Hern\'{a}ndez-Monteagudo for suggesting to compare the signal to noise ratios for different 
contemporary galaxy surveys.

\bibliography{bibtex/aamnem,bibtex/references}
\bibliographystyle{mn2e}

\appendix

\bsp

\label{lastpage}

\end{document}